\documentclass[fleqn,10pt]{wlscirep}
\usepackage[utf8]{inputenc}
\usepackage[T1]{fontenc}
\usepackage{lineno}
\usepackage{amsmath}

\linenumbers
\title{A large-scale multimodal dataset of human speech recognition}

\author[1]{Yao Ge}

\author[1,3]{Chong Tang}

\author[2]{Haobo Li}

\author[1]{Zikang Chen}

\author[4]{Wenda Li}

\author[3]{Kevin Chetty}
\author[2]{Daniele Faccio}

\author[1,*]{Qammer H. Abbasi}

\author[1]{Muhammad Imran}

\affil[1]{James Watt School of Engineering, University of Glasgow, Glasgow, G12 8QQ,UK}

\affil[2]{School of Physics \& Astronomy, University of Glasgow, Glasgow, G12 8QQ,UK}

\affil[3]{Department of Security and Crime Science, University College London, London, WC1E 6BT, UK}

\affil[4]{School of Science and Engineering, University of Dundee, Dundee, DD1 4HN, UK}

\affil[*]{corresponding author(s): Qammer H. Abbasi (qammer.abbasi@glasgow.ac.uk)}

\usepackage{subfig}
\usepackage{color}
\usepackage{booktabs}
\usepackage{caption}
\usepackage{multicol,multirow}

\begin{abstract}

Nowadays, non-privacy small-scale motion detection has attracted an increasing amount of research in remote sensing in speech recognition. These new modalities are employed to enhance and restore speech information from speakers of multiple types of data. In this paper, we propose a dataset contains 7.5 GHz Channel Impulse Response (CIR) data from ultra-wideband (UWB) radars, 77-GHz frequency modulated continuous wave (FMCW) data from millimetre wave (mmWave) radar, and laser data. Meanwhile, a depth camera is adopted to record the landmarks of the subject's lip and voice. Approximately 400 minutes of annotated speech profiles are provided, which are collected from 20 participants speaking 5 vowels, 15 words and 16 sentences. The dataset has been validated and has potential for the research of lip reading and multimodal speech recognition.
\end{abstract}

\begin{document}
\nolinenumbers
\flushbottom
\maketitle
\thispagestyle{empty}



\maketitle
\section{Background \& Summary}\label{sec1}
In general speech recognition tasks, acoustic information from microphones is the main source for analysing the verbal communication of human beings \cite{acoustic_review}. The speech process is not just a means of conveying linguistic information, which can also provide valuable insight into the speaker's characteristics such as gender, age, social and regional origin, health, emotional state, and in some cases even their identity. Recently, the automatic speech recognition (ASR) technique has already matured and been marketed \cite{asrreview}. In addition to sound signals, the series of physiological processes that produce sound, such as lip movement, vocal cord vibration, and head movement, also retain semantic and speaker information to some extent. On the other hand, there are two main limitations in specific environments that only audio information can not perfectly work for ASR: silent speech recognition (SSR) and multiple speakers environments. Both issues can be solved if considering the speaker physics properties.

First, SSR can be considered a significant branch of speech recognition that provides understandable and enhancing communication methods to assist patients with severe speech disorders. Recently, the directions of SSR is mainly focused on wearable sensors, which detect brain and muscle activity with electroencephalogram (EEG) sensor, articulator movements headset. and other kinds of implantable sensors \cite{ssi_review}. However, these methods are highly dependent on wearable and implant sensors, which is dedicated to patients but does not collect a large dataset from a normal person. Meanwhile, users should consider the potential health risk of contactable devices. For voice disorder and other patients who maintain the capability to control the vibration of vocal folds and face muscles, non-invasive SSR has the potential to improve their quality of life compared to electronic sensors.

In addition, in scenarios with multiple speakers, the microphone captures the sounds from the surroundings without distinguishing the person's identity, which seriously lowers the accuracy of speech recognition. This issue is similar to the cocktail party effect \cite{EEG_cocktail}, which is a phenomenon in which an individual can focus on one conversation despite being surrounded by several other simultaneous conversations. The effect is mainly attributed to the brain's ability to process auditory frequency and highlight certain sounds, allowing the individual to focus in on the source of interest without being easily distracted. However, it is a challenge to separate different sources only using acoustic data. In this case, additional radar or laser devices can assist the model in distinguishing the audio according to the physical information. For example, the proposed work \cite{mmwaveenhance1} combined the audio and radar signals to filter after added noise. And Secondly, the voice information including tone and speaking habits of individual contains a variety of personal data that can be used to create a unique voice fingerprint, such as speaking habits and intonation. This will cause a risk of sensitive data leakage, as the voice fingerprint could be used for identification. For the wireless sensing side-based algorithm, vocal folds vibration only focuses on the tone of speech, which does not include privacy information. Third, previous speech recognition research has focused mainly on visual-based mouth movements, posing a risk of lack of privacy and overlooking internal mouth movements.

\begin{figure}[ht]%
\centering
\includegraphics[width=0.6\textwidth]{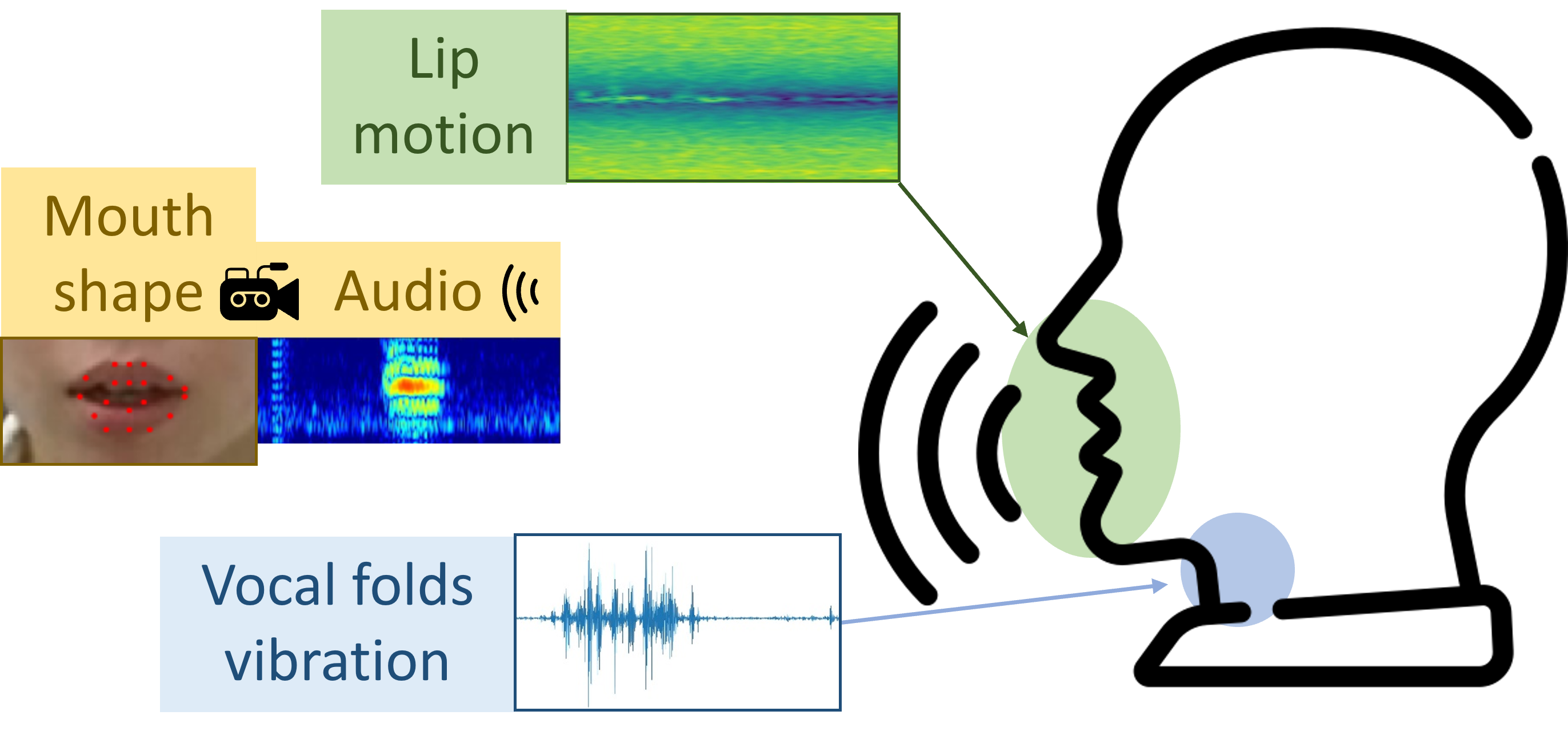}
\caption{Schematic diagram of the Multimodal signals }\label{fig0}
\end{figure}

In this paper, we proposed a dataset of human speech by collecting data from multiple sensors information while people are speaking specific corpus. The contributions of our dataset are concluded in following points:

\begin{enumerate}

\item To the best of our knowledge, we firstly proposed a modality of silent speech recognition and collect 6 hours dataset including the information from FMCW and UWB radar, laser, audio, video of mouth motion and processed mouth skeleton points. The dataset is expected to reduce the labour for researchers who expect to work on SSR or speech enhancement.
\item Our system takes into account the physical movements of all parts of the head during human speech, including mouth movements and vibrations of the vocal cord, which is illustrated in Fig. \ref{fig0}.
\item The diverse range of modalities in our dataset offers ample opportunities for conducting research in the field of speech recognition. The range contains but is not limited to the following application: radar-based vowels and words classification, speaker identification, speech enhancement in noisy environment, radar-based lip reconstruction, etc. 
\end{enumerate}

\section{Methods}\label{sec2}
Firstly, we conducted a literature review to establish the necessary sensors and experimental setup for radar-based speech recognition, given the absence of a standard and corpus. Meanwhile, we demonstrate the availability of all the sensors we adopted and then establish our data collection approach, referencing previous work. 

\paragraph{Literature survey of radar-enabled speech recognition}

There are various kinds of sensors has been adopted for speech research: UWB, mmWave radar and laser speckle detector\cite{radioses2022}. The work of UWB demonstrated the lip reading work with the vowels of A, E, I, O, U and static scenario, with a face mask. The result of 95\% approves that the mouth motion brings informative signal for UWB sensing \cite{2022pushing}. To expand the work and exploit more possibilities, we added words and sentences for data collection regarding the reference. Besides, mmWave FMCW radar has been used for speech enhancement in \cite{radioses2022,mmwaveenhance1}. These two researches has distinct focus directions: \cite{mmwaveenhance1} considered the distance coefficient for radar signal and successfully make speech enhancement system implementation in 7 meters; and \cite{radioses2022,vad2022} works on audio separation of multiple speakers with radar based spacial information. For laser-related information, \cite{cester2022remote} proposed a remote measurement technique for healthy individuals that involves capturing the reflected laser speckle from the surface of the neck skin. This system is capable of capturing the micro-vibration on the surface of the neck produced by blood pressure, which can also be adopted for the extraction of voice signals without audio signals through detecting the vibration caused by throat. Therefore, we adopted these devices as the core in our experiments.

\begin{figure}[ht]%
\centering
\includegraphics[width=0.9\textwidth]{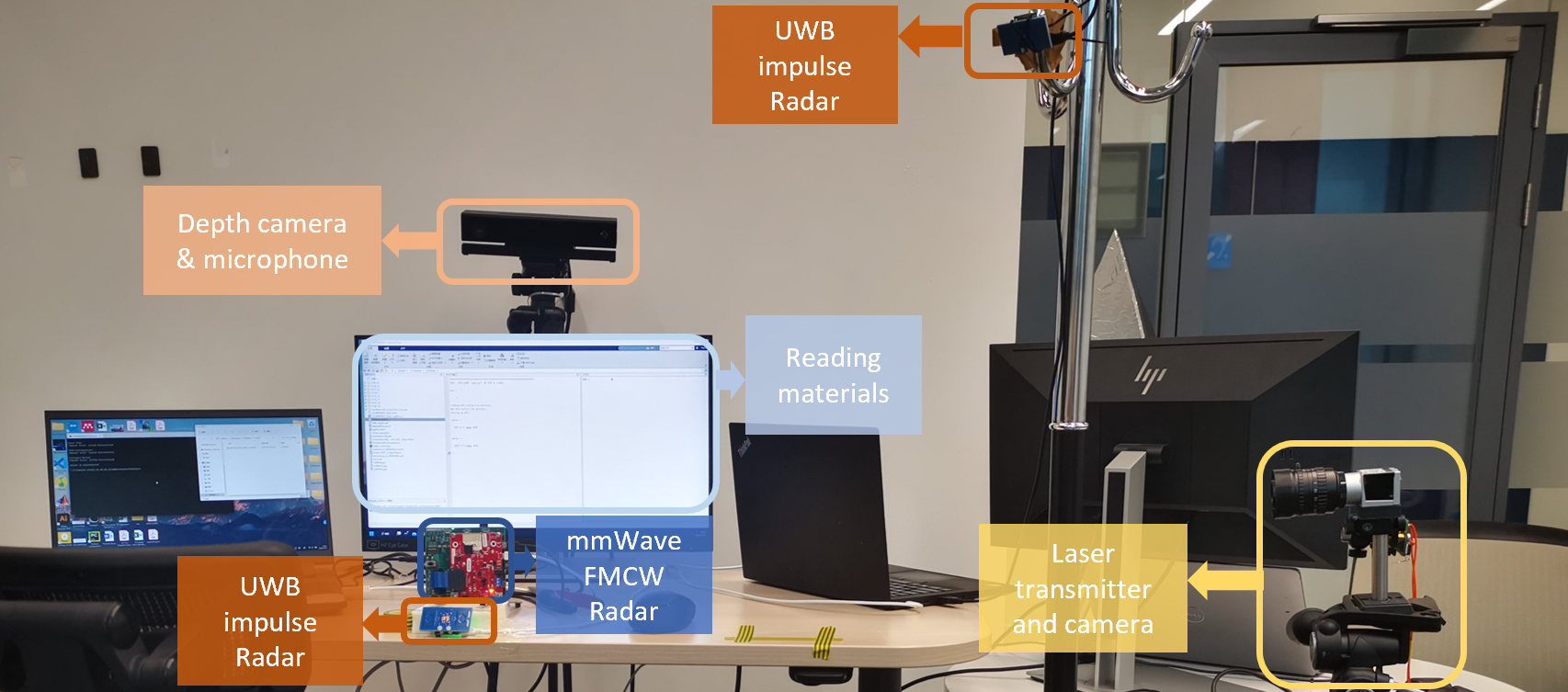}
\caption{Data collection setup with device label in the real scenario}\label{fig1}
\end{figure}

\paragraph{Data acquisition scheme}
The overall data collection system was organised by four laptops and four types of sensors: Microsoft Kinect V2 for audio and video including mouth landmark, X4M03 UWB radar from NOVELDA, AWR2243 mmWave radar from Texas Instrument, and laser measurement system for physical vibration of human speech. To keep the data synchronised with different sensors, we utilised TCP/IP connection to control the distinct host laptops with the same network time protocol (NTP) for recording the time stamp while data collection. Once we run script on master laptop, the master will send the commends to other three sockets in series. The mean delay from master to sockets of other devices is around 80 ms, which is considered in our post synchronisation processing.   
Considering the potential research for speech recognition, we designed three data collection schemes shown in the following. The adopted corpus is recorded in an additional folder in our dataset.
\begin{itemize}
    \item Single person speech of vowels, words and sentences.
    \item Dual-person speech simultaneously of complex sentences.
    \item Single person speech of vowels, words and sentences with different distance from radar to speakers.
\end{itemize}
The details of data collection from specific sensor are demonstrated below, with experiment setup shown in Fig. \ref{fig1}.

\paragraph{Speech voice}
We used Kinect v2 for collecting vocalised speech. Kinect v2 contains a 4-mic Microphone array. With the enable Kinect v2 to collect the accurate acoustic information. The sample rate of audio data is 16000Hz, bit depth is 16-bit. The frequency range of recording audio is up to 8000Hz, that can cover the frequency range of human voice.
\paragraph{Mouth skeletal points}
The Kinect v2 is also used in collecting the facial landmark information. A RGB camera and an infrared camera are intergrated in kinect v2. By measuring the time of flight (ToF) using the IR camera Kinect can get the depth image. Meanwhile, we use the lip recognition method proposed in \cite{haliassos2021lips} for extraction of the lip skeleton, which is provided as part of our dataset.
\paragraph{IR-UWB radar}
Like Wi-Fi and Bluetooth, UWB is a short-range wireless communication protocol. The UWB was defined as the wireless transmission system of which bandwidth exceeds 500 MHz, and each transmit pulse of this communication system can occupy at least 500 MHz bandwidth.
Instead of modulating with a carrier wave, IR-UWB relies on nanosecond (ns) to picosecond (ps) non-sinusoidal narrow impulse radio signals to transmit data. The time-based modulation technology increases transmission speed and reduces power consumption. For speech recognition, the UWB system has the following advantages:
\begin{enumerate}
\item Strong anti-interference ability: From the RF mechanism, the pulse wave emitted by UWB is more resistant to interference than the continuous electromagnetic waves in short range. Specifically, the permitted work frequency band of UWB is from 3 GHz to 10 GHz, which suffers less disturbance from general 2.4 GHz WiFi system and other telecommunication signals.
\item The protocol yielded positive results, resulting in a reduction in power consumption for short-range communication applications. The transmit power of the UWB transmitters was found to be typically less than 1 mW, which extended the system's operating time and minimised electromagnetic wave radiation to the human body.
\end{enumerate}

After careful consideration of cost and feasibility, we have selected the XeThru X4M03, an IR-UWB radar system on chip, as our UWB radar. The UWB RF specifications of this radar have been approved by ETSI (European Telecommunications Standards Institute) in Europe, and FCC (Federal Communications Commission) in the USA for commercial use in human living circumstances \cite{xethru_intro}.This device is a highly reliable sensor that is capable of detecting objects at a range of up to 10 meters. It is also capable of detecting objects in a wide range of angles, up to 180 degrees. This radar system has been adopted in a variety of research projects, ranging from human vital sign detection \cite{uwbradar1,uwbradar2} to activity recognition \cite{uwbact}.


For pulsed radars, the range between radar and target can be calculated by $R=\frac{c*\Delta T}{2}$, where $c$ represents microwave speed, $\Delta T$ represents the round-trip time of a single pulse, called time of flight (ToF). The signal of IR-UWB can be represented in Eq. \ref{Eq1}, where the $\tau$ represents the ToF of signal impulses in fast-time range, $t$ represents receiving time of frame in slow-time domain, $N_d$ is the index of the dynamic path, $a_i(\tau,t)$ represents the complex attenuation factor of the $i^{th}$ path; $e^{-j2\pi \frac{d(t)}{\lambda}}$ represents the phase change of $i^{th}$ path; $d_{i}(t)$ and $d_{a}(t)$ are the static length and vibrating length of $i^{th}$ path. $\lambda$ represents the wavelength of the UWB signal.   
\begin{equation} 
\textbf{s}(\tau,t) = \sum_{i=1}^{N_d}a_i(\tau,t)e^{-j2\pi \frac{(d_{i}(t) + d_{a}(\tau))}{\lambda}}
\label{Eq1}
\end{equation}

\begin{figure}[ht]%
\centering
\includegraphics[width=0.5\textwidth]{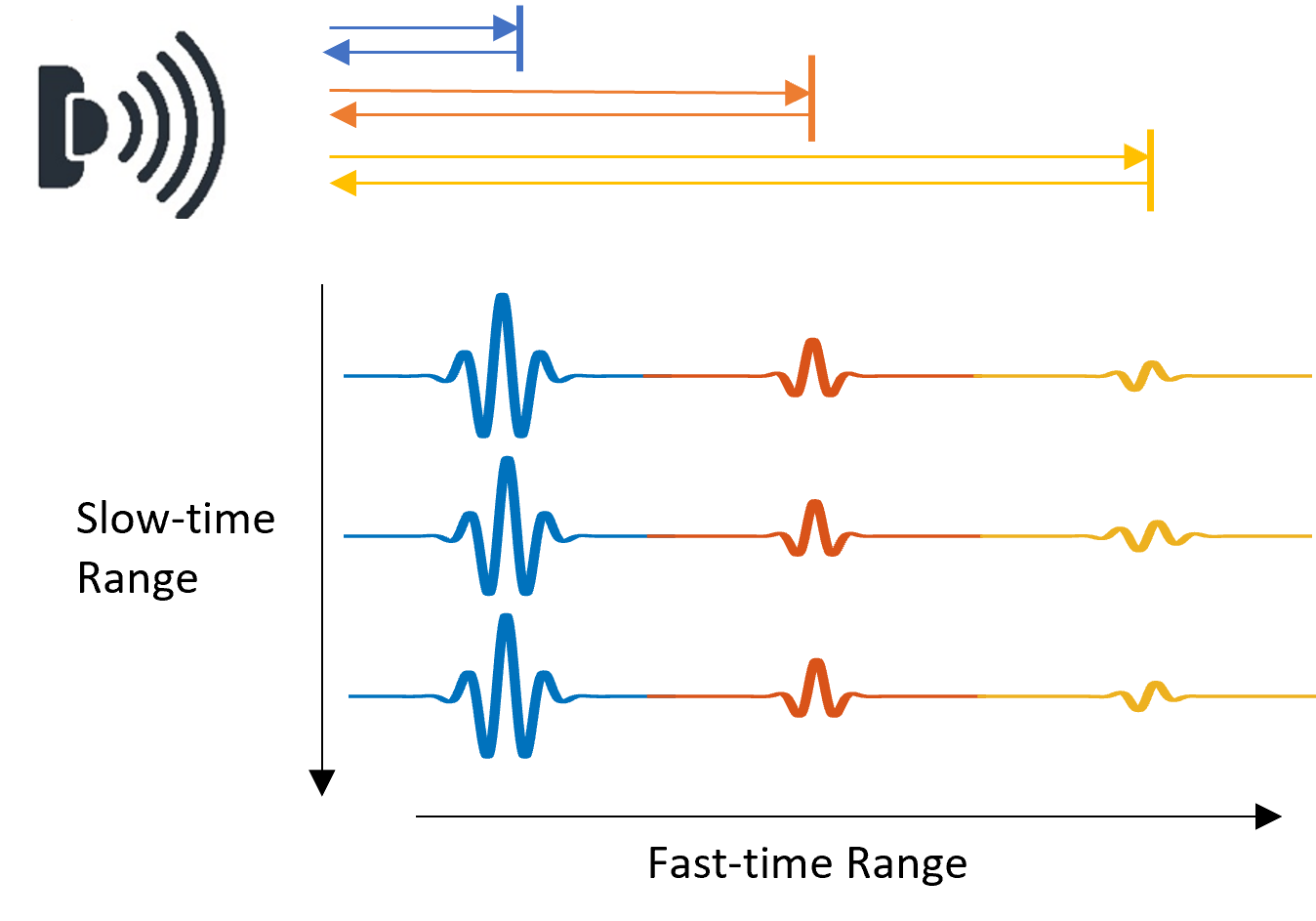}
\caption{IR-UWB impulse signal format}\label{fig2}
\end{figure}

The format of the received signals is a fixed set of bins determined by the of the transmitting pulses indexed by fast-time and slow-time shown in Fig. \ref{fig2}. Fast-time and slow-time are two dimensions that are used to describe the format of the received signals in a UWB radar system. Fast-time is the time it takes for the radar to transmit a pulse and receive the reflected signal. Slow-time is the time it takes for the radar to transmit multiple pulses and receive the reflected signals. The fast-time and slow-time dimensions are used to determine the format of the received signals, which is a fixed set of bins. These bins are used to store the information about the objects detected by the radar. The fast-time and slow-time dimensions are also used to determine the range of objects detected by the radar. By using the fast-time and slow-time dimensions, the UWB radar system can accurately detect objects in a wide range of distances. This makes the UWB radar system an ideal choice for a variety of applications, including human vital sign detection and activity recognition.

For data collection, we adopted Moduleconnector API which is supported by the radar on Windows MATLAB 2021b. The detailed parameters of radar is listed in the Table \ref{tab:subtab1}.

\begin{figure}[ht]%
\centering
\includegraphics[width=0.7\textwidth]{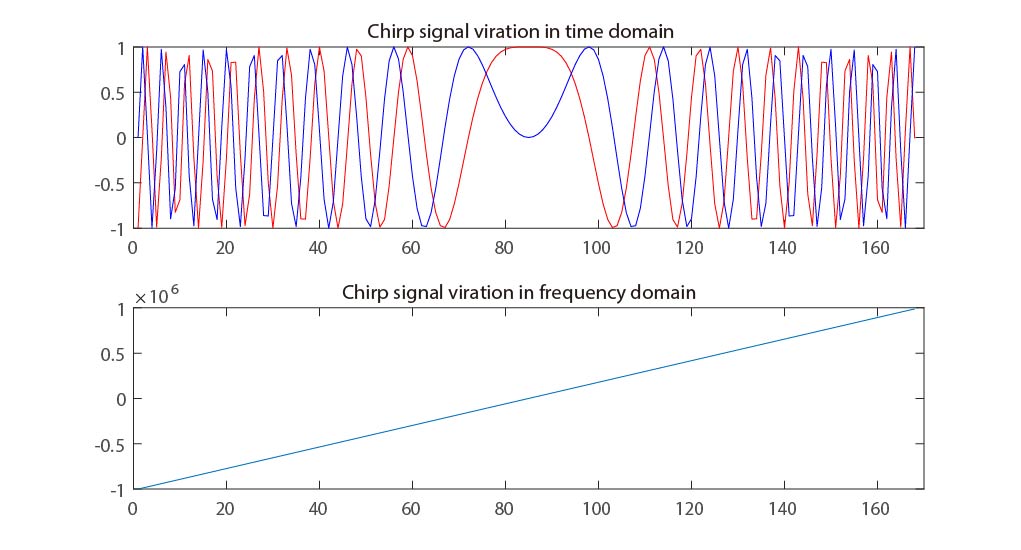}
\caption{The plot of mmWave FMCW radar chirp signal in time and frequency domain}\label{fig3}
\end{figure}

\paragraph{mmWave FMCW radar}
Although the IR-UWB radar is able to capture the vibration of sub-centimeter motion. The angle resolution is limited by the number of antennas. Meanwhile, for comparison of speech recognition performance using different modulation-based radars, we added one commercial off-the-shelf (COTS) 77 GHz FMCW radar, AWR2243, for data collection. This high frequency enables radar signal to capture motion in millimeter, that can be used for both lip motion and vocal folds detection. FMCW radar signal can be illustrated as Fig. \ref{fig3}.

Unlike IR-UWB radars that measure distance using the ToF of instantaneous impulses, FMCW radars rely on the difference in frequency between the transmitted and received signal from a linear variation of signal frequency. In other words, Doppler effect of moving target can be explored from difference between transmission frequency ($f_t$) and shifted frequency ($f_s$). The formula shows Doppler velocity of target relative to radar: $v=\frac{c(f_s-f_t)}{2ft}$.  

Except for modulation methods, AWR2243 radar contains 4 receive antennas and 3 transmit antennas, which is possible to adopt angle of arrival (AoA) for research in horizontal plane. In our experiment, we adopt 4 receive and 1 transmit antennas to increase the sample rate.
In data collection, we utilised mmWaveStudio API on Windows with the configuration file. The parameters were set to the value shown in Table \ref{tab:dev_subtab2}. 


\begin{table}[t]
    \centering
    \label{Tab:uwbsetting}
    \subfloat[X4M03 setup]{
        \centering
        \label{tab:subtab1}
        \begin{tabular}{l|l}
        \hline
        Parameter          & Value         \\ \hline
        Center frequency   & 8.745 GHz     \\
        Sampling frequency & 23.328 GHz    \\
        Frame rate         & 300 Hz        \\
        Bandwidth          & 1.5 GHz       \\
        Number of antennas & 1 Tx and 1 Rx \\ \hline
        \end{tabular}
    }
    \hspace{1cm}
    \subfloat[AWR2243 setup]{
        \centering
        \label{tab:dev_subtab2}
        \begin{tabular}{l|l}
        \hline
        Parameter              & Value         \\ \hline
        Center frequency       & 8.745 GHz     \\
        Sampling frequency     & 23 GHz        \\
        Frame rate             & 1018 Hz       \\
        Bandwidth              & 4GHz         \\
        Number of antennas     & 1 Tx and 4 Rx \\
        No.of Sample per chirp & 512           \\
        No. of Chirp per frame & 1             \\
        \hline
        \end{tabular}}
    \caption{Recognition accuracy vs User identity}

\end{table}

\paragraph{Laser}
The laser measurement system consists of a 532nm green laser diode (DJ532-40, Thorlabs) as transmitter and a high-speed CMOS camera from Basler as receiver, where the laser diode emits a laser beam pointing to the face outline of the testing subject and the camera captures the reflected laser speckle patterns. Both transmitter and receiver are fixed on a 1.2m tripod, and the camera is connected with a laptop via an USB 3.0 cable for powering and data transferring. The green laser diode has a distance of approximately 1 m to the participants, it will produce an illumination spot of around 5mm diameter on the human skin by considering the beam divergence. For the laser safety, laser power exposed on human skin is controlled to be less than 0.5 mw (CLASS 1), therefore it is safe for long-term eye and skin exposure. The focal length and f-stop of the camera objective are set as 25 mm and 0.95, respectively, allowing the camera system to detect the laser speckle from a very close range (0.1 m) to a relative far range (up to 3 m). Additionally, the size of the region of interest (ROI) window is chosen as 128x128 pixels, and the camera exposure time is set as 600 µs. The laser and camera are carefully aligned before the experiments to ensure that the selected ROI includes the movements of speckles. 
For every single measurement, the collected data is in a format of $W\times H\times N$, where W and H is represents the width and height of the ROI, respectively, and the N equals to the number of frames within the measuring period. In our case, N is correlated with the sampling frequency of the CMOS camera, which is set as 1.47 kHz.

\begin{figure*}[htbp]
\centering

\subfloat[Top View]{
    \label{figCon:subfig2}
    \includegraphics[width=0.3\linewidth,height=0.23\textheight]{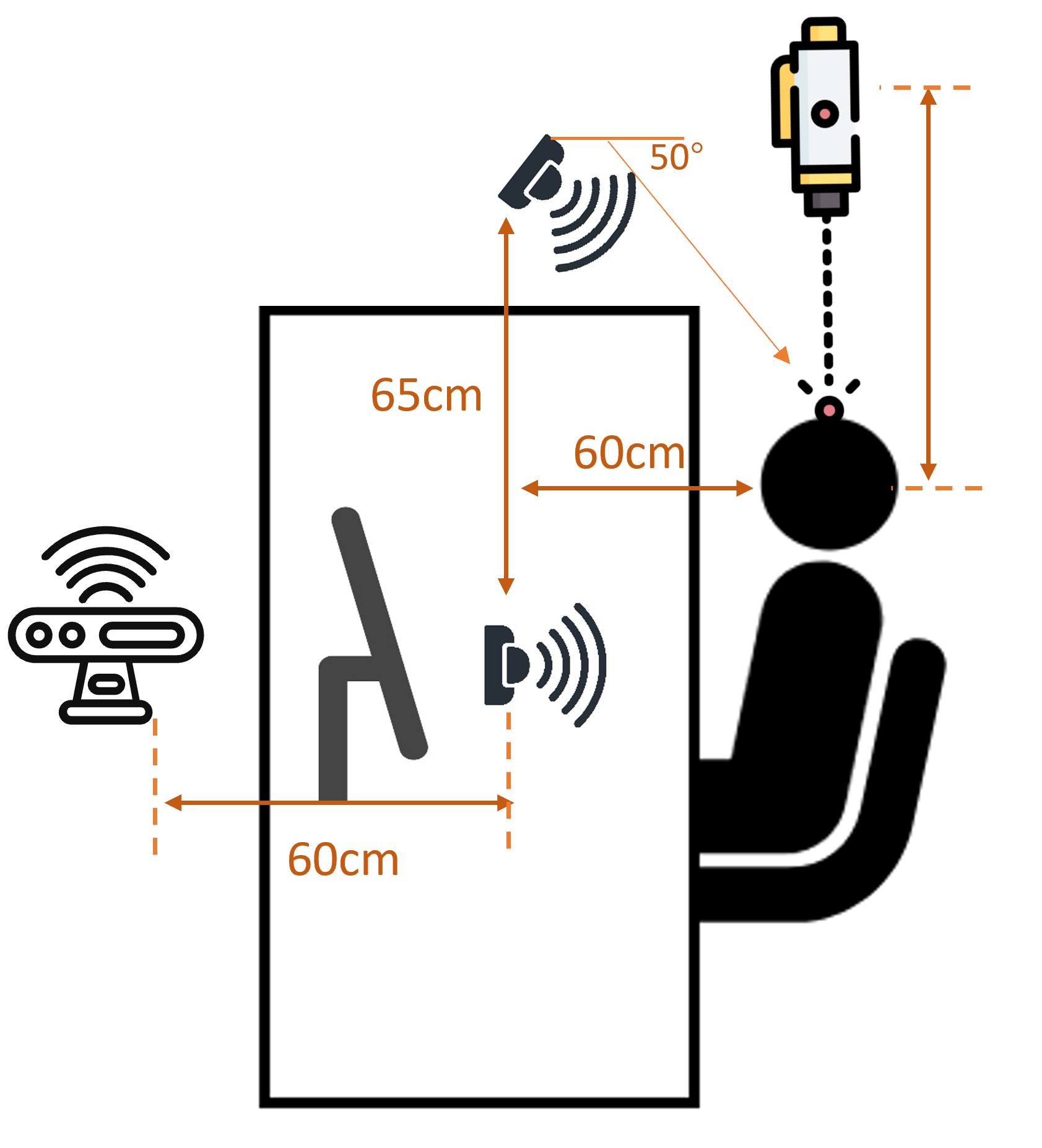}
}
\hspace{1cm}
\subfloat[Front View]{
    \label{figCon:subfig3}
    \includegraphics[width=0.45\linewidth,height=0.23\textheight]{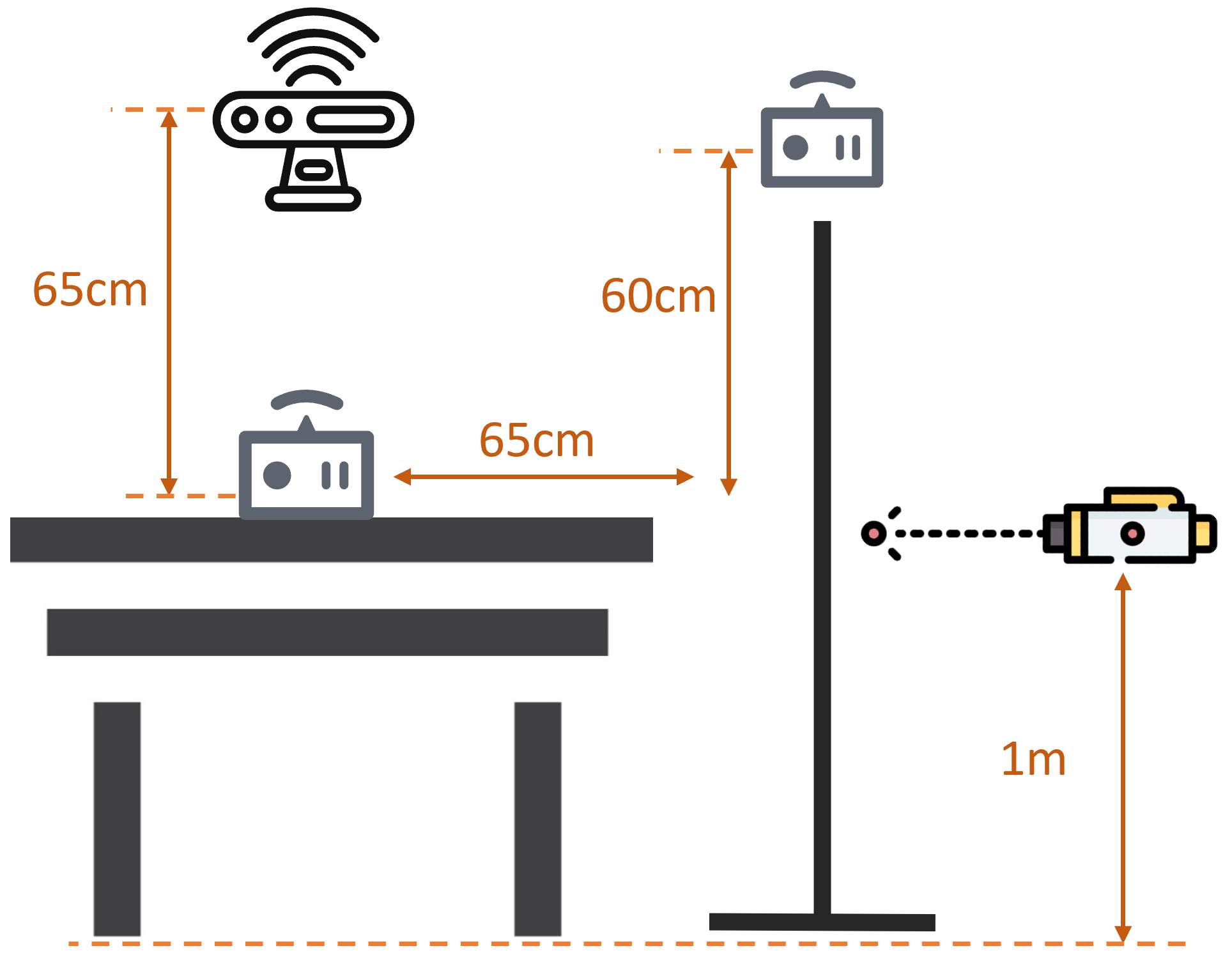}
}
\caption{Detailed setup schematic diagram for single person scenario (Laser's location is not fixed due to the camera based signal process only require the laser directly point to skin of subjects. The UWB radar facing to subject directly was called 'xe2' in dataset folder, another is called 'xe1')}
\label{fig4}
\end{figure*}

\paragraph{Participant}
There are 20 volunteers contributing to our experiment, whom come from different country regions including Europe, China, Pakistan and UK. Our dataset for speech recognition presents both opportunities for generalization and challenges due to the volunteers' diverse backgrounds, resulting in distinct accents. To the issue bring from body size, we adopted an adjustable table for subjects that can keep the relative distance same between the head of speaker and different sensors. The feature of accents and speech habits can be extracted from our dataset with lip motion, vocal folds vibration, and audio, which is potential for related multimodal ASR research.

Meanwhile, all participants were informed about the purpose of the study and what was expected of them. Experiment consent forms were obtained from each participant prior to the experiments. Whole project of dataset collection was ethical approved by University of Glasgow College of science and engineering (approval no: 300210309).

\section{Data Record}\label{sec3}
The multimodal speech detection dataset is accessible with the link of \href{https://nextcloud.gla.ac.uk/s/LJHKyBxLHXdk4xZ}{https://nextcloud.gla.ac.uk/s/LJHKyBxLHXdk4xZ}. The corpus of the single person scenario is listed in Table \ref{tab:2}. During data collection, we asked volunteers to pronounce a specific vowel / word / sentence with timestamps on laptops. All laptops were synchronised using the same NTP server. During collection, volunteers were guided by automatic voice instruction to read the corpus and relax. The timestamps of audio instruction were instantaneously recoreded in kinect/timestamp. However, there is a few seconds of uncontrolled latency in activating all radars and laser equipment, which disrupted the devices' ability to synchronise acquisition data. In this case, we decided to keep these devices recording for one minute and write timestamp while data collection is activated so that the signal can be cropped according to kinect timestamps. All were recorded alongside the data, which reduces the effort for manually separating the data. 

Meanwhile, we also collected data via different distances in the single-person scenario and the two-person scenario, with corpus listed in Table \ref{tab:3}. Instead of original $60cm$, we asked volunteers to sit $1.2m$ and $2.2m$ away from radar equipment, respectively, which is a potential for researchers to explore the relationship of radar-based audio detection with distance. Besides, in two-person experiment, we kept one volunteer sitting in the same place as the single-person scenario shown in Fig. \ref{fig4}, then let another speaker sit on the left side of the first-mentioned volunteer. The two subjects were asked to normally read different corpus shown in Table \ref{tab:4}, which was shown on the screen in one minute, without repeating words. The laser equipment was pointed to the first volunteer, and the kinect camera only took information from another subject. This kind of dataset will contribute to multiple audio source separation.

\begin{table}
\centering
\small
\begin{tabular}{|c|l|c|c|}
\hline
\multicolumn{1}{|l|}{\textbf{Type}} & \textbf{Corpus}                                                                                                                                                       & \multicolumn{1}{l|}{\textbf{Index}}                         & \multicolumn{1}{l|}{\textbf{Participants}} \\ \hline
vowel                               & a, e, i, o, u                                                                                                                                                         & \begin{tabular}[c]{@{}c@{}}1-5 \\ in sequence\end{tabular}  & User 1 - 20                                \\ \hline
word                                & \begin{tabular}[c]{@{}l@{}}order, assist, help, ambulance, bleed, \\ fall, shock, medical, sanitize, doctor,\\ accident, rescue, emergency, heart, break\end{tabular} & \begin{tabular}[c]{@{}c@{}}1-15 \\ in sequence\end{tabular} & User 1 - 20                                \\ \hline
\multirow{16}{*}{sentences}         & I need help.                                                                                                                                                          & 1                                                           & \multirow{7}{*}{User 1 - 20}               \\ \cline{2-3}
                                    & Call for an ambulance.                                                                                                                                                & 2                                                           &                                            \\ \cline{2-3}
                                    & The building's on fire.                                                                                                                                               & 3                                                           &                                            \\ \cline{2-3}
                                    & Can you smell smoke.                                                                                                                                                  & 4                                                           &                                            \\ \cline{2-3}
                                    & Where's the fire escape.                                                                                                                                              & 5                                                           &                                            \\ \cline{2-3}
                                    & There's been an accident.                                                                                                                                             & 6                                                           &                                            \\ \cline{2-3}
                                    & Is there a doctor here?                                                                                                                                               & 8                                                           &                                            \\ \cline{2-4} 
                                    & The staff sanitized the sickroom.                                                                                                                                     & 7                                                           & \multirow{3}{*}{User 1,2,4,6,7}            \\ \cline{2-3}
                                    & Medical care is important.                                                                                                                                            & 9                                                           &                                            \\ \cline{2-3}
                                    & Don't worry about bleeding.                                                                                                                                           & 10                                                          &                                            \\ \cline{2-4} 
                                    & I am having trouble breathing.                                                                                                                                        & 7                                                           & \multirow{3}{*}{User 3, 8 - 13}            \\ \cline{2-3}
                                    & I think I'm having a heart attack.                                                                                                                                    & 9                                                           &                                            \\ \cline{2-3}
                                    & My heart is failling.                                                                                                                                                 & 10                                                          &                                            \\ \cline{2-4} 
                                    & Need emergency treatment at shock stage.                                                                                                                              & 7                                                           & \multirow{3}{*}{User 5, 14 - 20}           \\ \cline{2-3}
                                    & He need a rescue for a heart attack.                                                                                                                                  & 9                                                           &                                            \\ \cline{2-3}
                                    & Don't worry about falling.                                                                                                                                            & 10                                                          &                                            \\ \hline

\end{tabular}
\caption{\label{tab:2}Corpus list for single subject experiment. The index of each participant is identical to the user label.}

\end{table}

\begin{table}
\small
\centering
\begin{tabular}{|c|l|c|l|}
\hline
\multicolumn{1}{|l|}{\textbf{Type}} & \textbf{Corpus}                                                                                                                                                                    & \multicolumn{1}{l|}{\textbf{Label}}                        & \textbf{Participants}                                                                                                                                                                                     \\ \hline
vowels                              & a, e, i, o, u                                                                                                                                                                      & \begin{tabular}[c]{@{}c@{}}v1-5 in \\ sequence\end{tabular} & \multicolumn{1}{c|}{\multirow{7}{*}{\begin{tabular}[c]{@{}c@{}}User 4 of 1.2m and 2.2m \\ (Index No. 24 and 25 in dataset), \\ User 5 of 1.2m and 2.2m \\ (Index No. 26 and 27 in dataset)\end{tabular}}} \\ \cline{1-3}
words                               & \begin{tabular}[c]{@{}l@{}}order, ambulance, medical,   \\ sanitize, accident\end{tabular}                                                                                         & \begin{tabular}[c]{@{}c@{}}w1-5 in \\ sequence\end{tabular} & \multicolumn{1}{c|}{}                                                                                                                                                                                     \\ \cline{1-3}
\multirow{5}{*}{sentences}          & Call for an ambulance                                                                                                                                                              & s1                                                          & \multicolumn{1}{c|}{}                                                                                                                                                                                     \\ \cline{2-3}
                                    & There's been an accident                                                                                                                                                           & s2                                                          & \multicolumn{1}{c|}{}                                                                                                                                                                                     \\ \cline{2-3}
                                    & The staff sanitized the sickroom                                                                                                                                                   & s3                                                          & \multicolumn{1}{c|}{}                                                                                                                                                                                     \\ \cline{2-3}
                                    & Is there a doctor here?                                                                                                                                                            & s4                                                          & \multicolumn{1}{c|}{}                                                                                                                                                                                     \\ \cline{2-3}
                                    & Medical care is important.                                                                                                                                                         & s5                                                          & \multicolumn{1}{c|}{}                                                                                                                                                                                     \\ \hline

\end{tabular}
\caption{\label{tab:3}Corpus list for supplementary experiments of changing the distance.}

\end{table}
\begin{table}
\begin{tabular}{|c|c|c|l|}
\hline
\multicolumn{1}{|l|}{\textbf{Type}} & \multicolumn{1}{l|}{\textbf{Corpus}}                                                                                                                                                & \multicolumn{1}{l|}{\textbf{Label}} & \textbf{Participants}                                  \\ \hline
\multirow{6}{*}{article}            & \multirow{6}{*}{\begin{tabular}[c]{@{}c@{}}From view of Kinect, volunteer \\ on the left side read 'Mr Sticky', \\ on the right side read \\ 'The king of the birds'.\end{tabular}} & b1-11                               & User 6 (Left) and User 4 (Right), recorded in Index 21 \\ \cline{3-4} 
                                    &                                                                                                                                                                                     & b12-22                              & User 4 (Left) and User 6 (Right), recorded in Index 21 \\ \cline{3-4} 
                                    &                                                                                                                                                                                     & b1-11                               & User 4 (Left) and User 5 (Right), recorded in Index 22 \\ \cline{3-4} 
                                    &                                                                                                                                                                                     & b12-23                              & User 5 (Left) and User 4 (Right), recorded in Index 22 \\ \cline{3-4} 
                                    &                                                                                                                                                                                     & b1-11                               & User 5 (Left) and User 1 (Right), recorded in Index 23 \\ \cline{3-4} 
                                    &                                                                                                                                                                                     & b12-23                              & User 1 (Left) and User 5 (Right), recorded in Index 23 \\ \hline
static                              & \multicolumn{1}{l|}{sitting without speaking}                                                                                                                                       & b24-26                              & User 5 (Left) and User 4 (Right), recorded in Index 22 \\ \hline
\end{tabular}
\caption{\label{tab:4}Corpus list for supplementary experiments of two-person scenario. The reading materials are referred from 
 corpus publication\cite{futrell2021natural}.}
\end{table}

\begin{figure}[htbp]%
\centering
\includegraphics[width=0.8\textwidth]{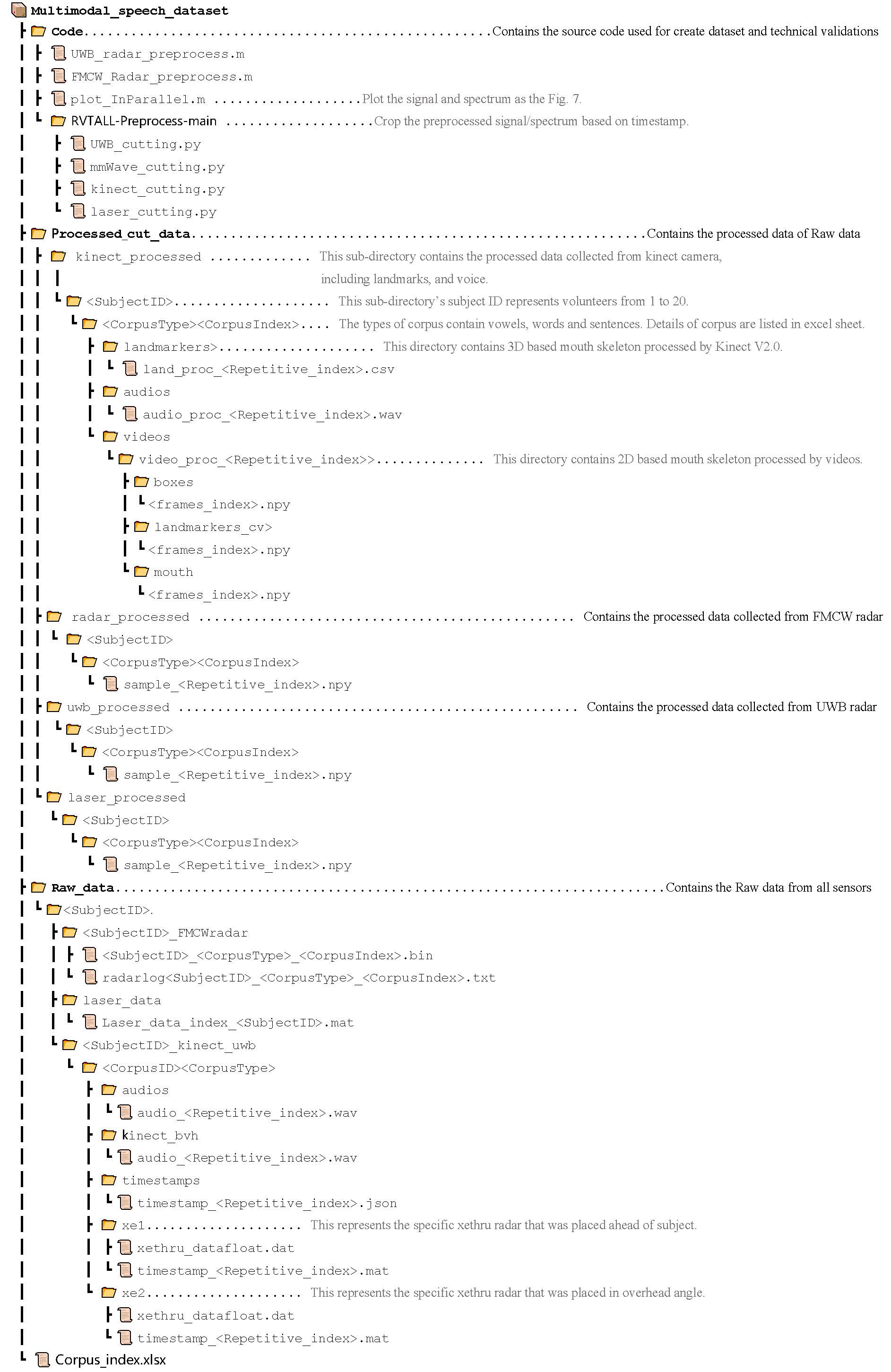}
\caption{The structure of the multimodal speech dataset}\label{fig5}
\end{figure}

\paragraph{Data storage structure}
After saving the data, all files were integrated into specific folders according to the data class, which is illustrated in Fig. \ref{fig5}. The entire dataset was divided into raw data and processed data to match the limitation of file size. Firstly, due to the data size limitation of file, we put mmWave FMCW radar data and laser data in a separate folder and other sensors in another. The radar signal files were kept in binary format with radar timestamps in text format. We provided the script for reading the timestamps along with radar signal in the code section demonstrated in Section \ref{sec6}. 
Meanwhile, information from kinect and two UWB radars was kept in same folder as the similar storage structure, which contains timestamps in JSON format, audio in WAV, landmarkers of user's head, and UWB radar signals in MAT format. 
Additionally, to ensure licence-free distribution of the dataset, the preprocessed data was converted to .npy, .csv and .wav files regarding the usages.

\section{Technical validation}\label{sec4}
\begin{figure}[htpb]%
\centering
\includegraphics[width=1\textwidth]{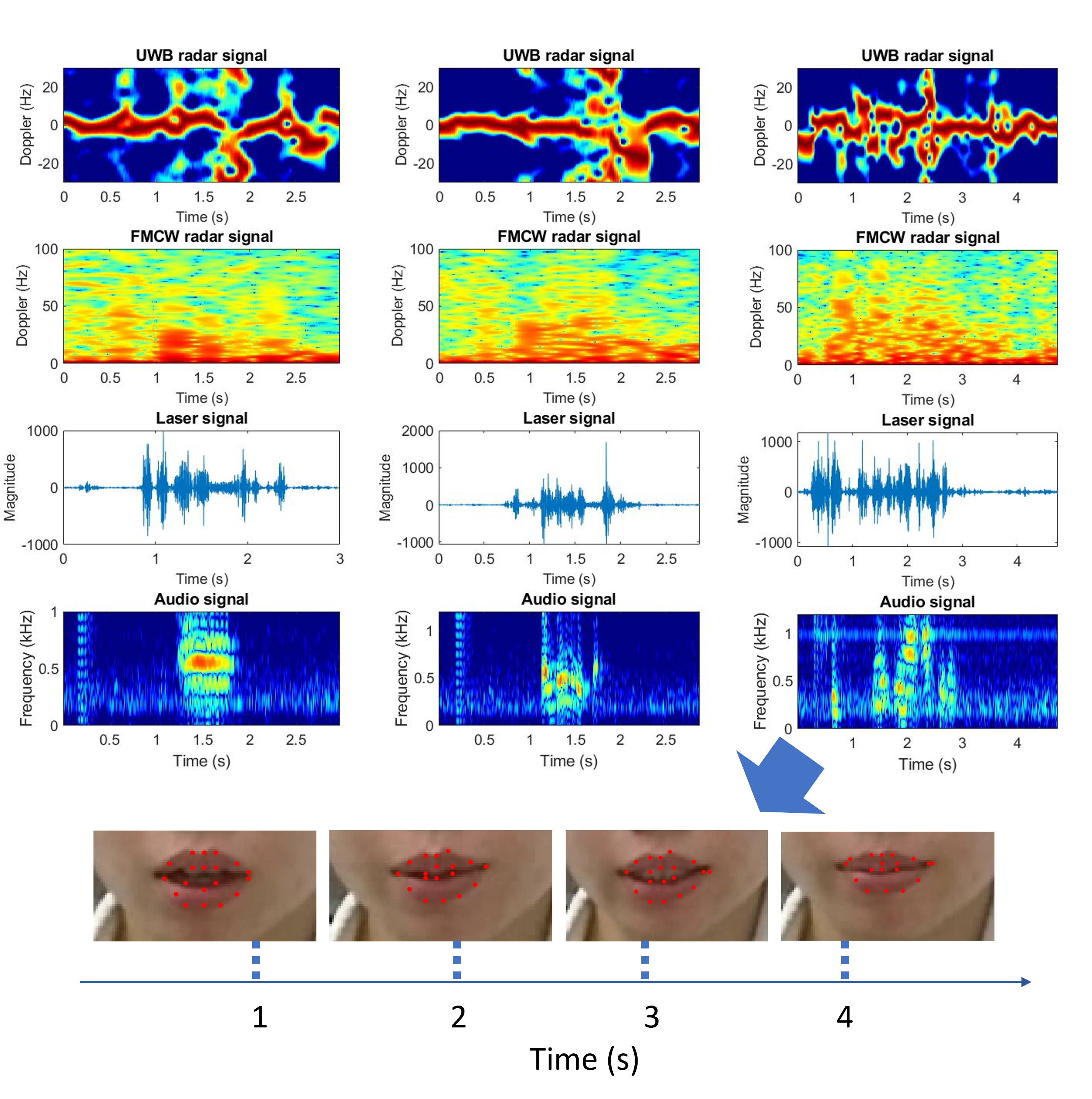}
\caption{Multimodal data illustration including UWB and mmWave radar signal, laser, audio, image, and mouth skeleton points. From left to right columns, the first represents the volunteer is speaking of vowel 'o', second is speaking of word 'bleed', and the third is speaking of sentence 'There's been an accident'. The last row illustrates the camera vision of the volunteer's mouth with the processed skeleton.}\label{fig6}
\end{figure}

The effectiveness of the collected data is validated in two parts for validation and benchmark: signal processing and analysis and UWB radar-based classification.
\paragraph{Signal analysis}
In this section, we analyse the entire process of lip motion and vibration of the vocal fold combined with video frames of the skeletal mouth and information about the voice, as shown in Fig. \ref{fig6}.

For UWB and FMCW radar signals, we transferred the raw data to the Doppler spectrum, shown with the speech spectrum and skeleton motion. 
The Fig. \ref{fig6}. shows all synchronised data types that were collected in the dataset. For UWB data, to sanitise the stationary object, the raw signal was first multiplied by a moving target indication (MTI) filter, which is a radar process method that allows the UWB radar to detect and track targets that are moving in relation to the radar devices. From Eq. \ref{Eq1}, we know that the channel impulses indicate different ranges. To consider all channel vibrations, we calculate the short-time Fourier transformation (STFT) result on each channel and then add all channels together, which is shown in Fig. \ref{fig6}.

For FMCW radar, we first transfer the IQ data to range-bin data through 1D-FFT. Then, instead of adopting the velocity dimension with multiple chirps, we increase the frame rate to obtain continuous and smooth radar signals. Because in our case, we are not interested in the velocity of mouth and vocal folds, but in the frequency response reflected by radar signals. The frequency range of vocal folds is from $80 Hz$ to $400 Hz$ \cite{vocalsfold}. Our radar's frame rate is adjusted to $1017 Hz$, which fits the minimum requirement of Nyquist criterion, which is the sampling rate should be more than twice of the highest frequency contained in the signal. Next, considering the wide range of radar that can be obtained, we introduce the FFT in the AoA dimension to obtain angle information regarding the location of the radar, which can be demonstrated in \cite{fmcw_aoa}. Meanwhile, MTI was applied to filter out the noise reflected by static object. We subtract the two spectrograms at a specific time interval which can reduce the occurrence of false alarms resulting from significant indoor clutter to a certain degree. Through getting the radar strength of the range of interest, which represents the location of the human, we get the information about the movement of the mouth and the vibration of the vocal folds.


In addition, we transfer the video to images with 30 frame per second and voice signal in spectrum, shown in Fig. \ref{fig6} together with Doppler spectrograms of UWB and FMCW radar signals. To retrieve the sound signal from raw laser data, an optical flow-based method, notably, the Farneback algorithm, is utilised to estimate the displacement of laser speckles on participants' faces. The input of this algorithm is every frame, denoted as a 2-D function \(f(x,y)\), whereas a quadratic polynomial expansion is adopted to approximate the grey value of each pixel and its neighbours. The signal model based on the local coordinates of the selected pixel could be written as the Eq. \ref{eq9}.
\begin{equation}
f(x)=x^{T}Mx+n^{T}+q
\label{eq9}
\end{equation} 
where x is the local coordinate \((x,y)\),\(M\) is a symmetric matrix equal to 
$
\begin{bmatrix}
 C_{4}& C_{6}/2\\ 
 C_{6}/2& C_{5}
 \end{bmatrix} 
$
, n is a vector equal to 
$
\begin{bmatrix}
 C_{2}\\ 
 C_{3}
\end{bmatrix} 
$
and q is a scalar equal to \(C_{1}\), \(C_{1}\) to \(C_{6}\) are the coefficients of the quadratic polynomial expansion. The new signal could be expressed using a displacement index \(\Delta d\) as the Eq. \ref{eq10} indicates.
\begin{equation}
f(x-\Delta d)=(x-\Delta d)^{T}M_{1}(x-\Delta d)+n_{1}^{T}(x-\Delta d)+q_{1}\\
g(x)=x^{T}M_{2}x+n_{2}^{T}+q_{2}
\label{eq10}
\end{equation} 
Simply let \(f(x-\Delta d)=g(x)\), then we can get \(n_{2}=n_{1}-2M_{1}\Delta d\), leading to the solution of displacement index. Then the computed optical flow needs to be filtered with a band-pass filter. The cut-off frequency of the filter is chosen to be 80 and 255 Hz for removing the frequency components caused by non-speaking activities such as head and skin movement. 
Then we integrate all sorts of cropped data that were mentioned above and observe that the speech features in different ranges are synchronised among all kinds of signals by intuition.   

\begin{figure}[htpb]%
\centering
\includegraphics[width=1\textwidth]{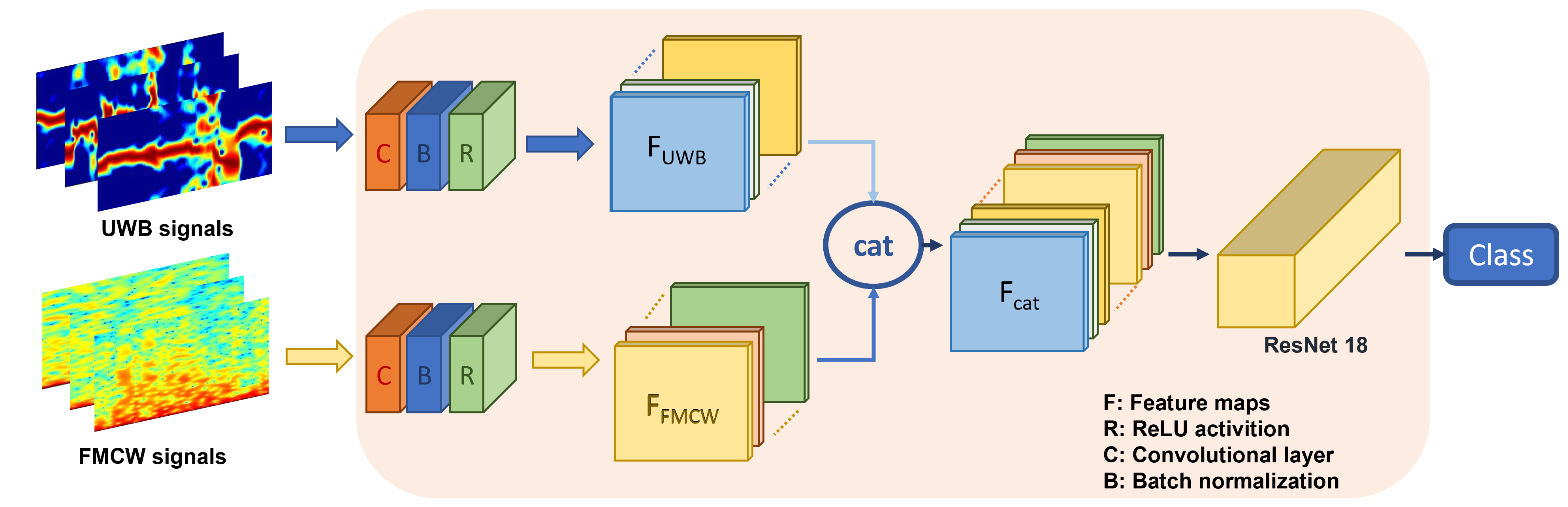}
\caption{Multi-modal sensor fusion scheme for word speech classification.}\label{fig8}
\end{figure}

\begin{figure}[htbp]
\centering

\subfloat[Confusion matrix of word classification based on UWB radar.]{
    \label{fig:cm1}
    \includegraphics[width=0.45\linewidth,height=0.26\textheight]{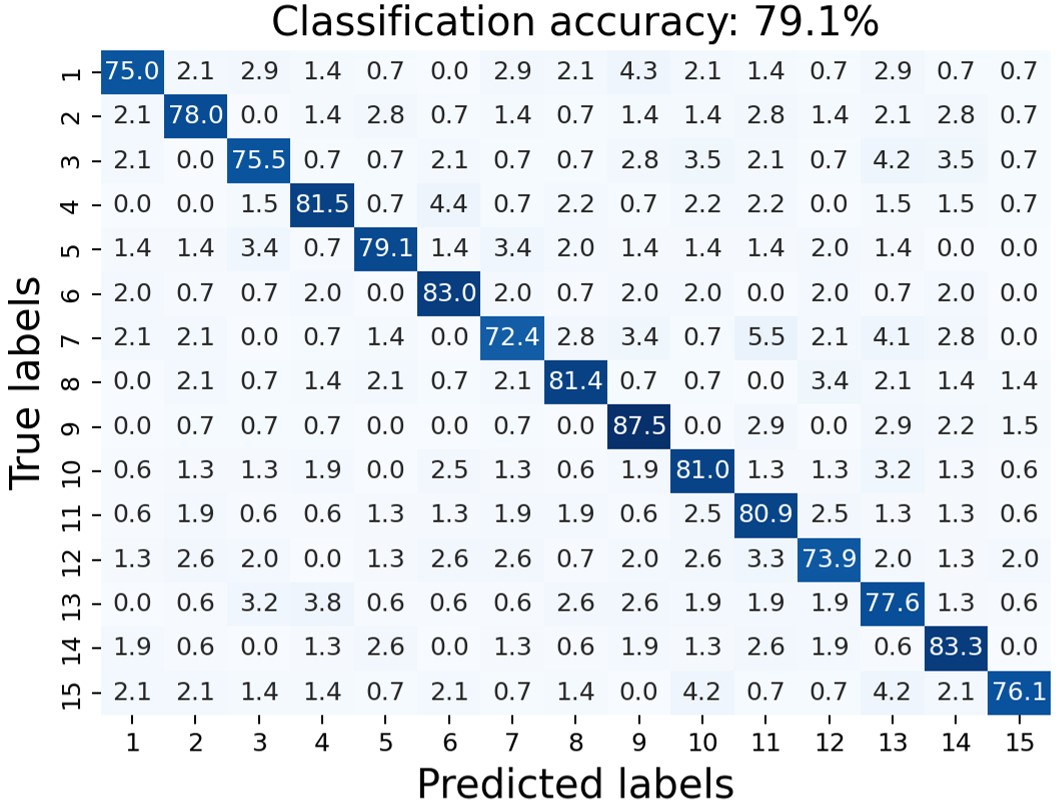}
}
\subfloat[Confusion matrix of word classification based on mmWave radar.]{
    \label{fig:cm2}
    \includegraphics[width=0.45\linewidth,height=0.26\textheight]{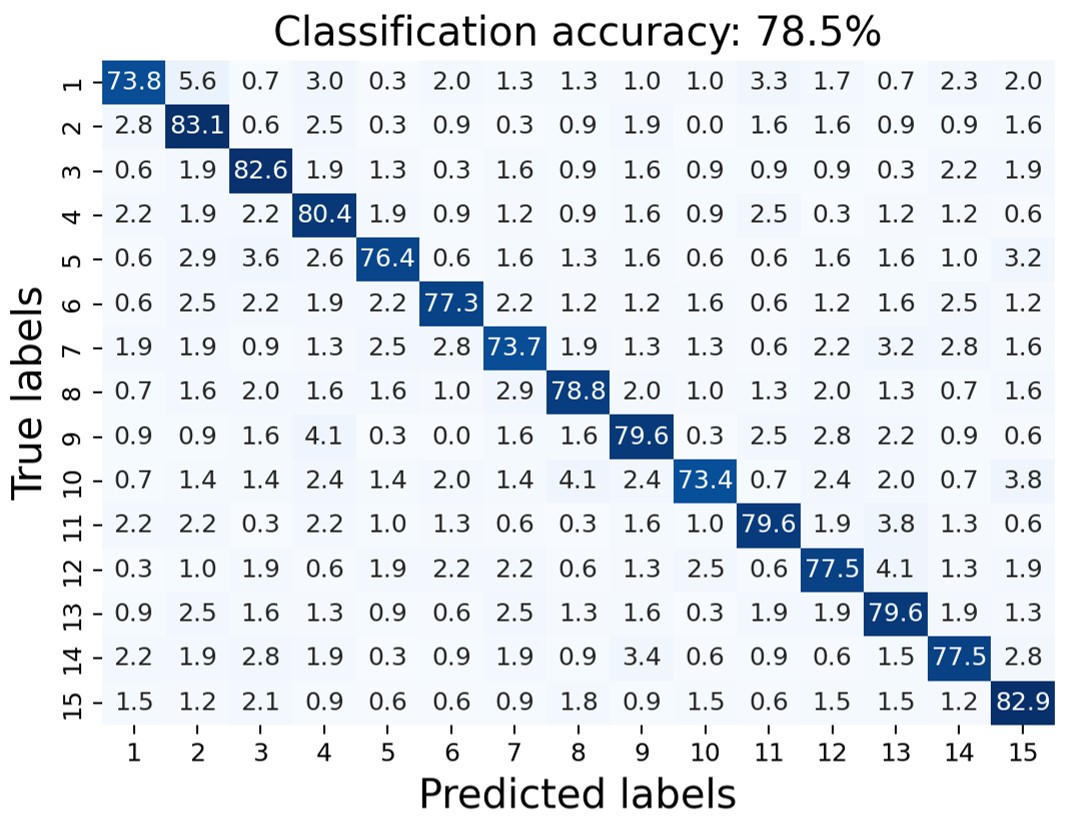}
}
\hspace{0.1cm}
\subfloat[Confusion matrix of word classification based on two modalities of UWB and mmWave radar.]{
    \label{fig:cm3}
    \includegraphics[width=0.45\linewidth,height=0.26\textheight]{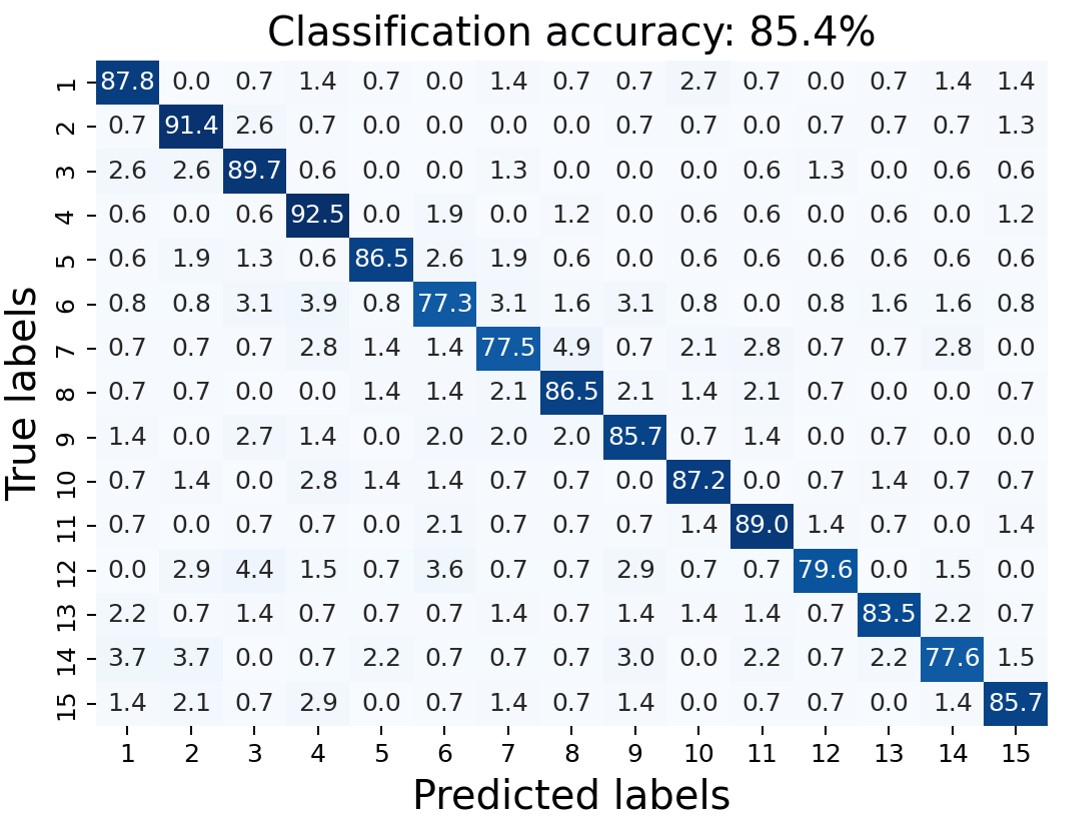}
}
\subfloat[Confusion matrix of word classification based on audio signal.]{
    \label{fig:cm4}
    \includegraphics[width=0.45\linewidth,height=0.26\textheight]{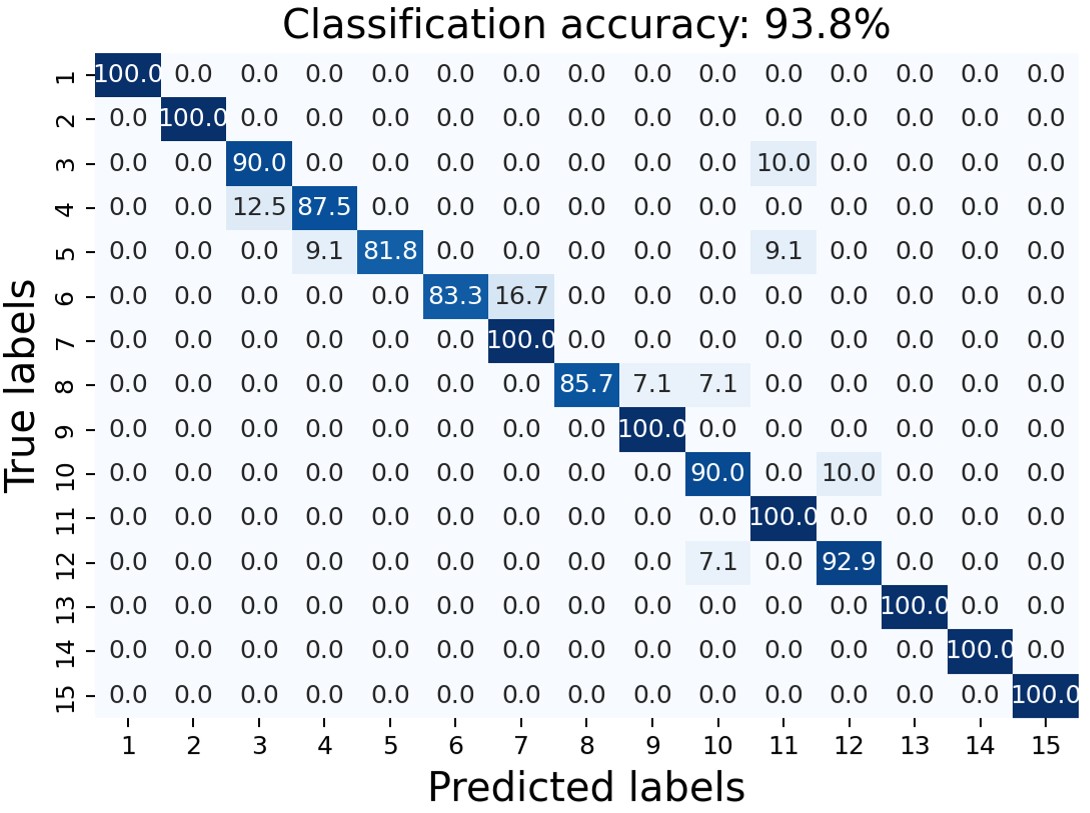}
}
\caption{Results of human speech classification task among 15 words labelled with UWB radar, mmWave radar, combination of two modalities of radars, and audio signals, respectively.}
\label{fig9}
\end{figure}

\paragraph{Multimodal words recognition}
For the benchmark of multimodal speech classification, we gathered UWB radar, mmWave radar and audio data of 20 subjects and established CNN-based ResNet classification network. In this case we only adopt one UWB radar that faces the person.
Meanwhile, we further consider a sensor fusion scheme that combines data from mmWave radar and UWB radar for radio-based word recognition. We employ a multi-input ResNet18 for this task, which includes two input blocks consisting of a convolutional layer, a batch normalisation layer, and a ReLU activation, as shown in Fig. \ref{fig8}. The initial feature extraction is completed by feeding spectrograms from the mmWave and UWB radar into their respective input blocks. The resulting feature maps are then stacked along the channel axis and processed by ResNet18 for final analysis. Then, we also applied audio data on word recognition as comparison with radio-based methods. The performance of the four mentioned systems are shown in Fig. \ref{fig9}. They illustrated that the combination of two radar modalities can assist system to get better results. It is also a challenge from the data fusion side that how to combine multi-modality information for different tasks.

\section{Usage Notes}
The reader is encouraged to analyse the data with the example script provided with the dataset. The functionalities of the script are described in the following section. Meanwhile, processing methods are not limited to the script we provide. We hope that readers can utilise the dataset regarding specific target. We would like to recommend that the user who wants to process their own algorithm on raw data, pay attention to the synchronisation among multiple sources with provided timestamps. Meanwhile, we created a github Repo for dataset code update of: \href{https://github.com/G-Bob/Multimodal-dataset-for-human-speech-recognition}{https://github.com/G-Bob/Multimodal-dataset-for-human-speech-recognition}. 

The video resources cannot be accessed in this dataset because they pertain to privacy concerns of vol. Please contact us by email to acquire further dataset with proper usage and data management licence.

\section{Code availability}
\label{sec6}
Matlab and Python scripts are provided in the codes directory of dataset for the users to replicate some of the figures:

\begin{itemize}
    \item[$-$] \textit{FMCW\_Radar\_process.m} This script is used to load the raw signal recorded by the AWR2243 radar. Then it is used to visualise the first and second FFT through distance dimension and angle dimension, respectively. Lastly, by reading the human location's phase variation, we can get human-related signals, including lip motion.
    \item[$-$] \textit{UWB\_radar\_process.m} This script is used to load the raw signal recorded by the Xethru X4M03 radar and process the data to STFT spectrums.
    \item[$-$] \textit{plot\_InParallel.m} This script provides a template to plot the spectrums that are shown on paper. First, the preprocessing data is needed to be downloaded.
    \item[$-$] \textit{uwb\_cutting.py; mmWave\_cutting.py; kinect\_cutting.py; laser\_cutting.py;} This script can be utilised to cut the sequence with given Kinect timestamp. In advance of this step, the radar signal should be processed to spectrums. This step can be used with the radar scripts we provide.  
\end{itemize}

\bibliography{sn-article}


\section*{Acknowledgements} 
This work was supported in parts by Engineering and Physical Sciences Research Council (EPSRC) grants EP/T021020/1 and EP/T021063/1 and by the RSE SAPHIRE grant.

\section*{Author contributions statement}
Y.G., C.T. and H.L. conceived the whole experiment, Q.A., D.F., K.C., W.L. and M.I. reviewed the experiment plan and supervised the whole experiment. H.L., C.T., Y.G. and Q.A. made the necessary applications and obtained the risk assessment and ethics approvals prior to the experiment. Y.G., H.L. and Z.C. conducted the experiment and recorded the experimental data from mentioned multiple modalities. C.T. and Y.G. analyzed the datasets. Y.G., C.T. and H.L. wrote the scripts in Matlab and Python for loading and analyzing the dataset from each modality. Y.G. lead to sort the time synchronisation and document experiments in this manuscript with the assistance by C.T., H.L. and Z.C. C.T. and Z.C. managed to write the classification network as part of validation. All authors reviewed the manuscript, provided constructive feedback and assisted with the editing of the manuscript prior to submission.

\section*{Competing interests}
The authors declare no competing interests.

\end{document}